\documentclass[ reprint, superscriptaddress]{revtex4-2}

\usepackage{natbib}
\usepackage{hyperref}
\usepackage[utf8]{inputenc}
\usepackage{amsmath}
\usepackage{tabularx}
\usepackage[bottom=3cm, top=2.7cm, left=2.7cm, right=2.7cm]{geometry}
\usepackage[dvipsnames]{xcolor}
\usepackage{subcaption}
\usepackage{graphicx} 
\usepackage{units} 
\usepackage[euler]{textgreek}
\usepackage{upgreek}
\graphicspath{ {data/} }

\newcommand{\ega}{e$^{-}$/$\gamma$~}

\begin{document}

\title{Particle discrimination in a NaI crystal using \\the COSINUS remote TES design}

\author{G.~Angloher}
\affiliation{Max-Planck-Institut f\"ur Physik, 80805 M\"unchen - Germany}

\author{M.R.~Bharadwaj}
\affiliation{Max-Planck-Institut f\"ur Physik, 80805 M\"unchen - Germany}

\author{I.~Dafinei}
\affiliation{Gran Sasso Science Institute, 67100 L'Aquila - Italy}
\affiliation{INFN - Sezione di Roma, 00185 Roma - Italy}

\author{N.~Di~Marco}
\affiliation{Gran Sasso Science Institute, 67100 L'Aquila - Italy}
\affiliation{INFN - Laboratori Nazionali del Gran Sasso, 67010 Assergi - Italy}

\author{L.~Einfalt}
\affiliation{Institut f\"ur Hochenergiephysik der \"Osterreichischen Akademie der Wissenschaften, 1050 Wien - Austria}
\affiliation{Atominstitut, Technische Universit\"at Wien, 1020 Wien - Austria}

\author{F.~Ferroni}
\affiliation{INFN - Sezione di Roma, 00185 Roma - Italy}
\affiliation{Gran Sasso Science Institute, 67100 L'Aquila - Italy}

\author{S.~Fichtinger}
\affiliation{Institut f\"ur Hochenergiephysik der \"Osterreichischen Akademie der Wissenschaften, 1050 Wien - Austria}

\author{A.~Filipponi}
\affiliation{Dipartimento di Scienze Fisiche e Chimiche, Universit\`a degli Studi dell'Aquila, 67100 L'Aquila - Italy}
\affiliation{INFN - Laboratori Nazionali del Gran Sasso, 67010 Assergi - Italy}

\author{T.~Frank}
\affiliation{Max-Planck-Institut f\"ur Physik, 80805 M\"unchen - Germany}

\author{M.~Friedl}
\affiliation{Institut f\"ur Hochenergiephysik der \"Osterreichischen Akademie der Wissenschaften, 1050 Wien - Austria}

\author{A.~Fuss}
\affiliation{Institut f\"ur Hochenergiephysik der \"Osterreichischen Akademie der Wissenschaften, 1050 Wien - Austria}
\affiliation{Atominstitut, Technische Universit\"at Wien, 1020 Wien - Austria}

\author{Z.~Ge}
\affiliation{SICCAS - Shanghai Institute of Ceramics, Shanghai - P.R.China 200050}

\author{M.~Heikinheimo}
\affiliation{Helsinki Institute of Physics, Univ. of Helsinki, 00014 Helsinki - Finland}

\author{M. N. Hughes~}
\affiliation{Max-Planck-Institut f\"ur Physik, 80805 M\"unchen - Germany}

\author{K.~Huitu}
\affiliation{Helsinki Institute of Physics, Univ. of Helsinki, 00014 Helsinki - Finland}

\author{M.~Kellermann}
\affiliation{Max-Planck-Institut f\"ur Physik, 80805 M\"unchen - Germany}

\author{R.~Maji}
\affiliation{Institut f\"ur Hochenergiephysik der \"Osterreichischen Akademie der Wissenschaften, 1050 Wien - Austria}
\affiliation{Atominstitut, Technische Universit\"at Wien, 1020 Wien - Austria}

\author{M.~Mancuso}
\affiliation{Max-Planck-Institut f\"ur Physik, 80805 M\"unchen - Germany}

\author{L.~Pagnanini}
\affiliation{Gran Sasso Science Institute, 67100 L'Aquila - Italy}
\affiliation{INFN - Laboratori Nazionali del Gran Sasso, 67010 Assergi - Italy}

\author{F.~Petricca}
\affiliation{Max-Planck-Institut f\"ur Physik, 80805 M\"unchen - Germany}

\author{S.~Pirro}
\affiliation{INFN - Laboratori Nazionali del Gran Sasso, 67010 Assergi - Italy}

\author{F.~Pr\"obst}
\affiliation{Max-Planck-Institut f\"ur Physik, 80805 M\"unchen - Germany}

\author{G.~Profeta}
\affiliation{Dipartimento di Scienze Fisiche e Chimiche, Universit\`a degli Studi dell'Aquila, 67100 L'Aquila - Italy}
\affiliation{INFN - Laboratori Nazionali del Gran Sasso, 67010 Assergi - Italy} 

\author{A.~Puiu}
\affiliation{Gran Sasso Science Institute, 67100 L'Aquila - Italy}
\affiliation{INFN - Laboratori Nazionali del Gran Sasso, 67010 Assergi - Italy}

\author{F.~Reindl}
\affiliation{Institut f\"ur Hochenergiephysik der \"Osterreichischen Akademie der Wissenschaften, 1050 Wien - Austria}
\affiliation{Atominstitut, Technische Universit\"at Wien, 1020 Wien - Austria}

\author{K.~Sch\"affner}
\affiliation{Max-Planck-Institut f\"ur Physik, 80805 M\"unchen - Germany}

\author{J.~Schieck}
\affiliation{Institut f\"ur Hochenergiephysik der \"Osterreichischen Akademie der Wissenschaften, 1050 Wien - Austria}
\affiliation{Atominstitut, Technische Universit\"at Wien, 1020 Wien - Austria}

\author{D.~Schmiedmayer}
\affiliation{Institut f\"ur Hochenergiephysik der \"Osterreichischen Akademie der Wissenschaften, 1050 Wien - Austria}
\affiliation{Atominstitut, Technische Universit\"at Wien, 1020 Wien - Austria}

\author{C.~Schwertner}
\affiliation{Institut f\"ur Hochenergiephysik der \"Osterreichischen Akademie der Wissenschaften, 1050 Wien - Austria}
\affiliation{Atominstitut, Technische Universit\"at Wien, 1020 Wien - Austria}

\author{M.~Stahlberg}
\email{Corresponding author, martin.stahlberg@mpp.mpg.de}
\affiliation{Max-Planck-Institut f\"ur Physik, 80805 M\"unchen - Germany}

\author{A.~Stendahl}
\affiliation{Helsinki Institute of Physics, Univ. of Helsinki, 00014 Helsinki - Finland}

\author{M.~Stukel}
\affiliation{Gran Sasso Science Institute, 67100 L'Aquila - Italy}
\affiliation{INFN - Laboratori Nazionali del Gran Sasso, 67010 Assergi - Italy}

\author{C.~Tresca}
\affiliation{Dipartimento di Scienze Fisiche e Chimiche, Universit\`a degli Studi dell'Aquila, 67100 L'Aquila - Italy}
\affiliation{INFN - Laboratori Nazionali del Gran Sasso, 67010 Assergi - Italy}

\author{F.~Wagner}
\affiliation{Institut f\"ur Hochenergiephysik der \"Osterreichischen Akademie der Wissenschaften, 1050 Wien - Austria}

\author{S.~Yue}
\affiliation{SICCAS - Shanghai Institute of Ceramics, Shanghai - P.R.China 200050}

\author{V.~Zema}
\email{Corresponding author, vanessa.zema@mpp.mpg.de}
\affiliation{Max-Planck-Institut f\"ur Physik, 80805 M\"unchen - Germany}

\author{Y.~Zhu}
\affiliation{SICCAS - Shanghai Institute of Ceramics, Shanghai - P.R.China 200050}

\collaboration{The COSINUS Collaboration}
\author{A.~Bento}
\affiliation{LIBPhys-UC, Physics Departments, University of Coimbra, 3004-516 Coimbra - Portugal}
\affiliation{Max-Planck-Institut f\"ur Physik, 80805 M\"unchen - Germany}
\author{L.~Canonica}
\affiliation{Max-Planck-Institut f\"ur Physik, 80805 M\"unchen - Germany}
\author{A.~Garai}
\affiliation{Max-Planck-Institut f\"ur Physik, 80805 M\"unchen - Germany}

\date{\today}

\begin{abstract}
The COSINUS direct dark matter experiment situated at Laboratori Nazionali del Gran Sasso in Italy is set to investigate the nature of the annually modulating signal detected by the DAMA/LIBRA experiment. COSINUS has already demonstrated that sodium iodide crystals can be operated at mK temperature as cryogenic scintillating calorimeters using transition edge sensors, despite the complication of handling a hygroscopic and low melting point material. With results from a new COSINUS prototype, we show that particle discrimination on an event-by-event basis in NaI is feasible using the dual-channel readout of both phonons and scintillation light. The detector was mounted in the novel \textit{remoTES} design and operated in an above-ground facility for \unit[9.06]{g$\cdot$d} of exposure. With a 3.7~g NaI crystal, \ega events could be clearly distinguished from nuclear recoils down to the nuclear recoil energy threshold of \unit[15]{keV}. 

\end{abstract}

\maketitle
\section{Introduction}
In the field of direct dark matter searches null-results have been reported by most experiments~\cite{billard2022direct} with the notable exception of DAMA/LIBRA~\cite{bernabei2022recent}. DAMA/LIBRA measures scintillation light created by particle interactions in NaI target crystals at room temperature. An annual modulation of the recorded event rate has been observed for many years, which is consistent with dark matter particle interactions, but incompatible with results from other direct searches in this interpretation~\cite{billard2022direct}. The origin of the signal remains unclear. Several experiments have set out to study this phenomenon using the same target material, and strong constraints on the modulation amplitude have been reported by COSINE-100 \cite{adhikari2022three} and ANAIS \cite{amare2022dark}, which follow a similar detection principle as DAMA/LIBRA. Among the NaI experiments, COSINUS (Cryogenic Observatory for SIgnals seen in Next generation Underground Searches) will be the only one to feature a direct measurement of the nuclear recoil energy per event. This is possible through the use of transition edge sensors (TES) which are coupled to the NaI target crystals to provide another channel in addition to the scintillation light. NaI poses certain difficulties when operated in this calorimetric approach, such as hygroscopicity and a low melting point~\cite{angloher2017results}. A solution to this problem is the remoTES design, where the TES sensor is deposited on a separate wafer, which is then coupled to the absorber crystal using an Au-wire and pad~\cite{angloher2023first}. The first results of this coupling scheme for detectors with Si and TeO$_2$ absorbers were described in Ref.~\cite{angloher2023first}. We demonstrate in this work that the same principle is applicable to NaI crystals, and present results from the first NaI-remoTES detector. In particular, we show that discrimination between \ega events and nuclear recoils on an event-by-event basis is possible in NaI, which constitutes a milestone for COSINUS.

\section{Detector module}
\label{sec:design}

\begin{figure}[ht]
\centering
   \includegraphics[width=\linewidth]{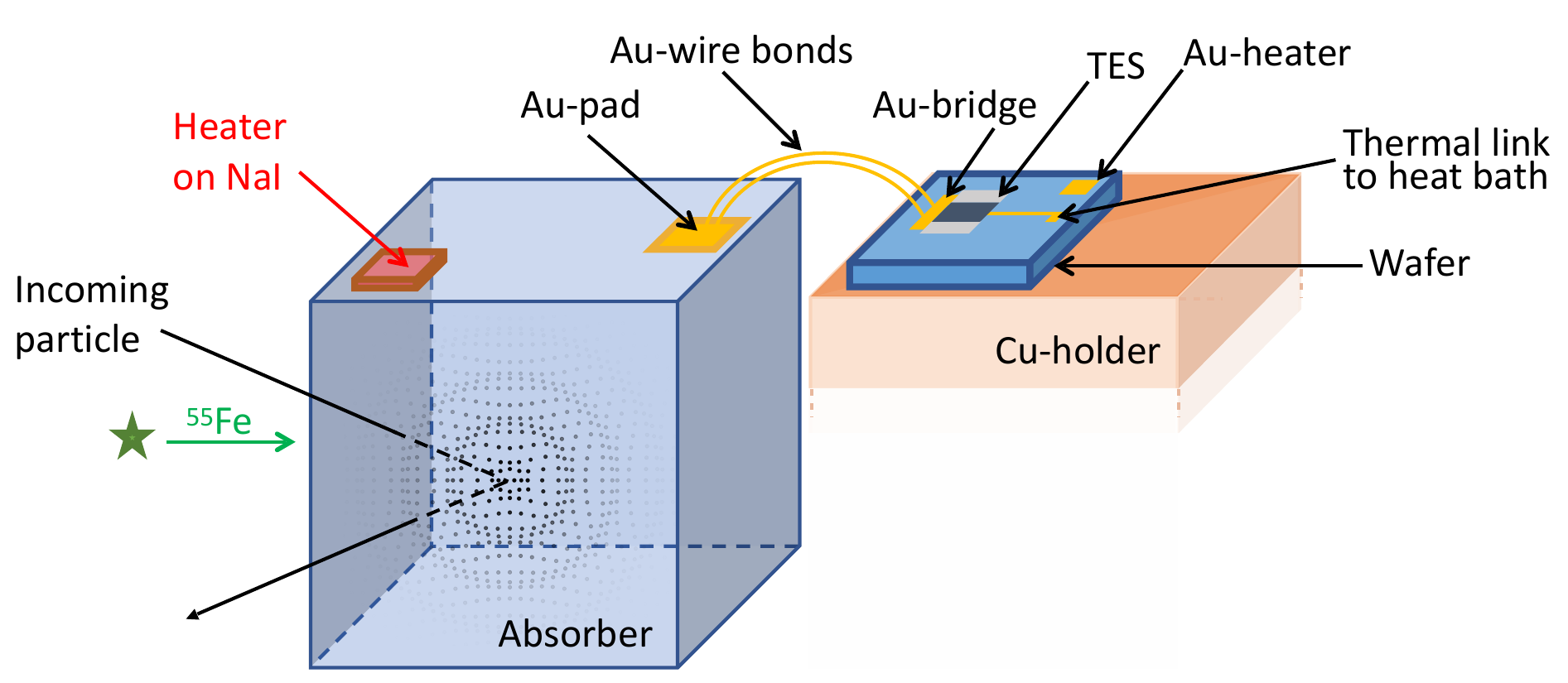}
   \caption{Schematic of the remoTES detector. The TES is deposited on a wafer, which is separated from the absorber crystal. The coupling between the absorber and the TES consists of an Au-pad glued on the absorber surface and connected to the TES by two Au-wire bonds~\cite{angloher2023first}.}
   \label{fig:scheme}
\end{figure}

\begin{figure*}[ht]
\begin{subfigure}[b]{0.47\textwidth}
 \centering
  \includegraphics[width=\textwidth]{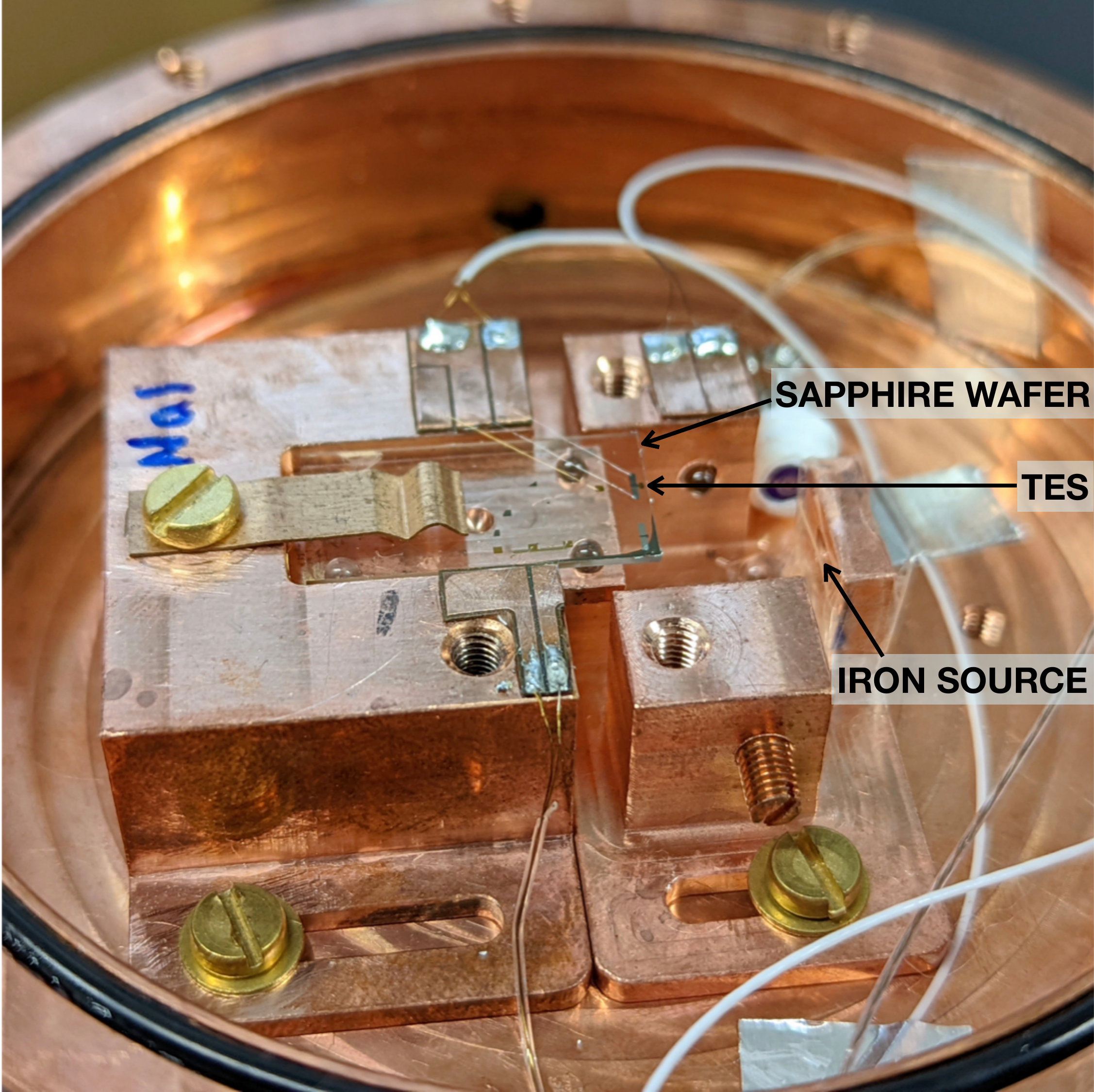}
   
  \caption{}
  \label{fig:wafer}
    \end{subfigure}
       \begin{subfigure}[b]{0.47\textwidth}
       \centering
     \includegraphics[width=0.995\textwidth]{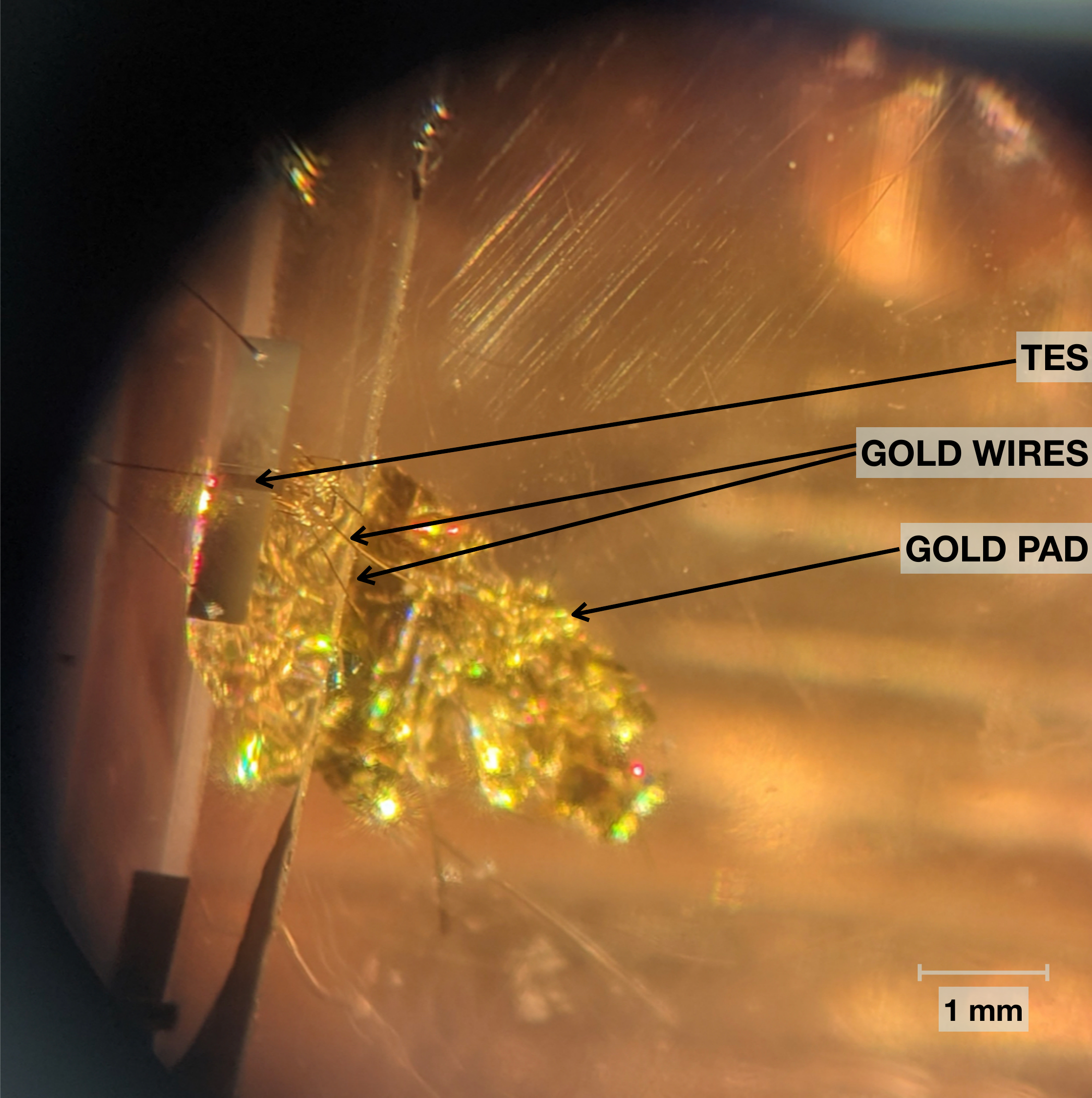}
        \caption{}
         \label{fig:pad}
   \end{subfigure}
  \caption{(a) The remoTES wafer mounted in the copper holder.  The sapphire wafer is placed on sapphire balls and held by a bronze clamp. A $^{55}$Fe source is taped on a copper piece facing the absorber for the purpose of energy calibration. (b) Microscopic view of the Au-pad glued on the NaI-crystal and the wire bonding to the remoTES.}
  \label{fig:remoTES}
\end{figure*}

\begin{figure*}[ht]
\begin{subfigure}[b]{0.47\textwidth}
 \centering
   \includegraphics[width=0.989\textwidth]{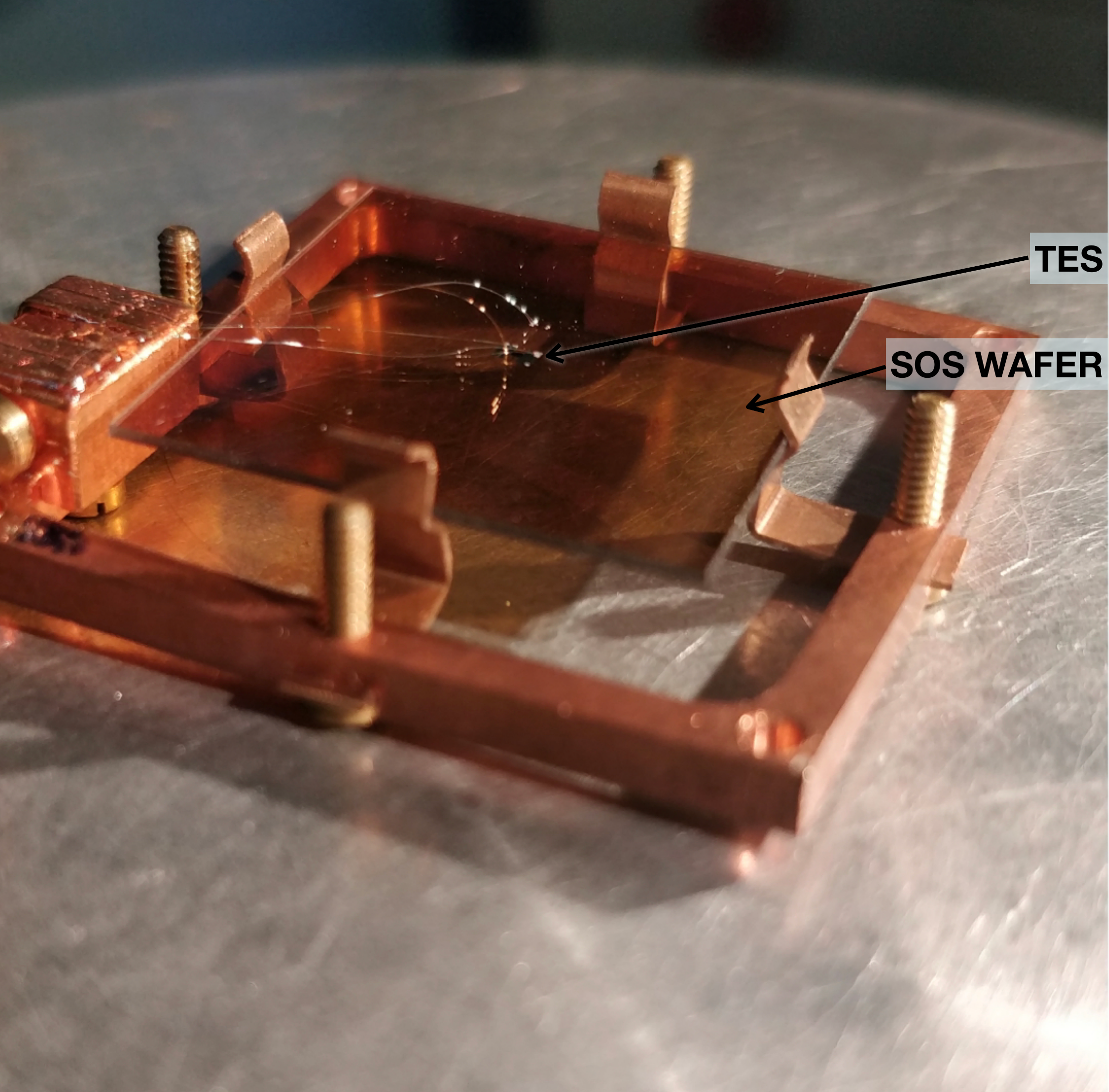}
  \caption{}
  \label{fig:LDholder}
    \end{subfigure}
       \begin{subfigure}[b]{0.47\textwidth}
       \centering
      \includegraphics[width=\textwidth]{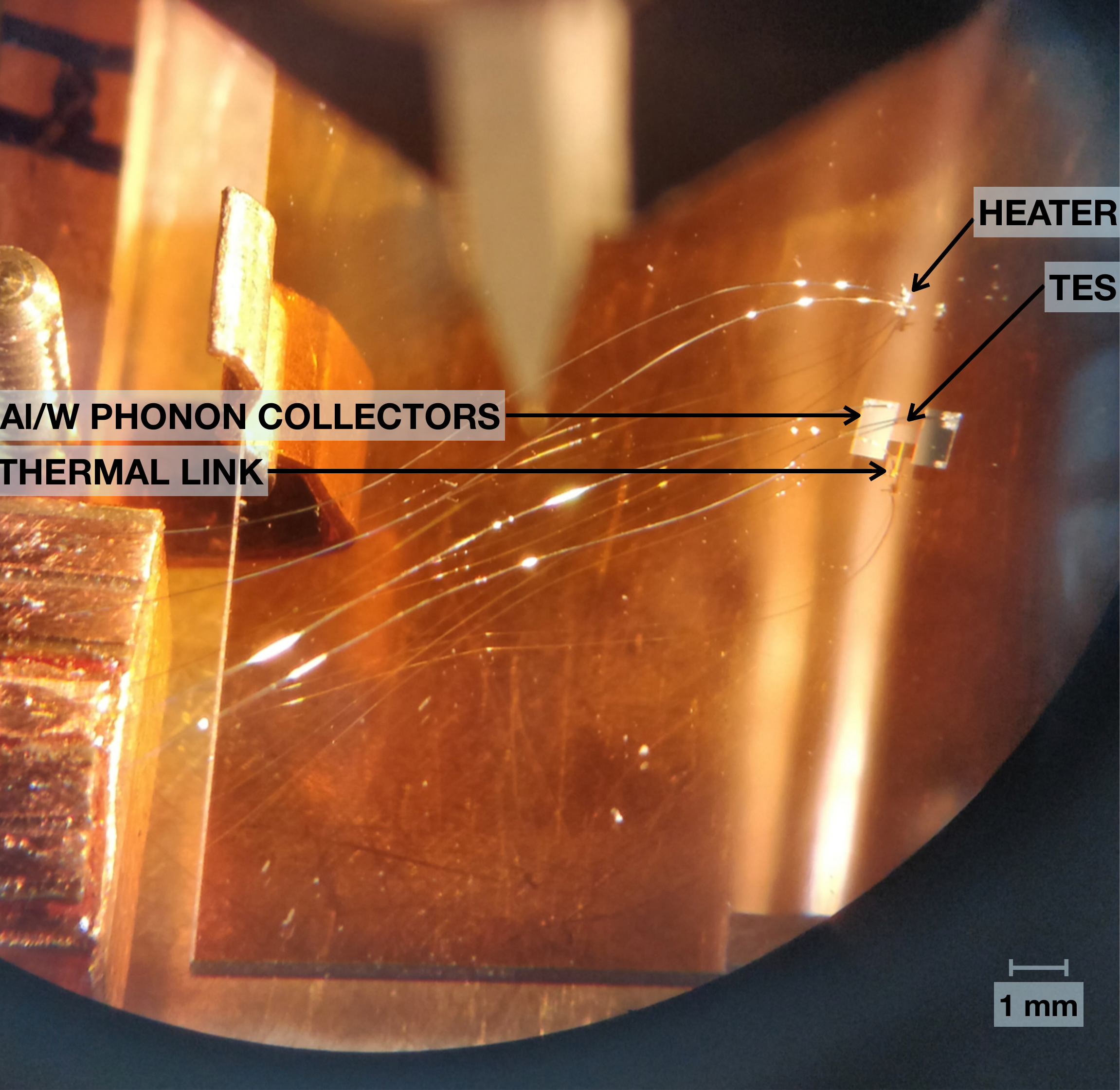}
        \caption{}
         \label{fig:LDtes}
   \end{subfigure}
  \caption{(a) SOS light detector mounted in a copper holder. (b) Microscopic view of the electrical connections of the light detector TES and of its separate ohmic heater.}
  \label{fig:LD}
\end{figure*}

The detector module consists of a remoTES phonon detector (cf.~\cite{angloher2023first}) shown in Fig.~\ref{fig:scheme} and \ref{fig:wafer} and a silicon-on-sapphire (SOS) wafer as light detector (cf. Fig.~\ref{fig:LD}).~The absorber is a (10$\times$10$\times$10)\,mm$^3$ NaI-crystal with a mass of 3.7\,g, manufactured by the Shanghai Institute of Ceramics (SICCAS). An Au-foil, cut from an ingot to a thickness of 1\,$\upmu$m and an area of 4\,mm$^2$, was glued on the absorber with EPO-TEK 301-2, a two component low out-gassing epoxy resin~\footnote{https://www.epotek.com/}. The residual resistivity ratio (RRR) of the Au-foil is about 22. The Au-foil was coupled to the TES wafer with two Au-wire bonds with a diameter of 17\,$\upmu$m each and lengths of 6.7\,mm and 10.3\,mm. A zoomed picture of the remoTES coupling to the absorber is shown in Fig.~\ref{fig:pad}.~An ohmic heater, fabricated on a (3$\times$3)\,mm$^2$ silicon pad with a thickness of 1~mm, was glued with EPO-TEK 301-2 on the surface of the NaI-absorber, and was used to inject heater pulses into the crystal. A $^{55}$Fe X-ray source with an activity of 3\,mBq was taped to the copper holder facing the NaI-absorber, and irradiated it from the side as indicated in Fig.~\ref{fig:wafer}. The wafer is a (10$\times$20$\times$1)\,mm$^3$ Al$_2$O$_3$ crystal, with an evaporated W-TES of (100$\times$407)~$\upmu$m$^2$ in area and a thickness of 156\,nm with two aluminum bonding pads for the connection to the bias circuit. The Au-wires from the Au-foil are bonded on the Au-bridge which overlaps with the W-film (see Fig.~\ref{fig:scheme}).~ Another Au-stripe is used as a thermal link connecting the TES to the thermal bath; its resistance is about 82.3~$\Omega$ at room temperature.~An Au-film with an area of (200$\times$150)~$\upmu$m$^2$ and a thickness of 100~nm, deposited on the wafer surface, is used to inject heater pulses and thus monitor the temperature of the TES.

The light detector is a (20$\times$20$\times$0.4)~mm$^3$ SOS wafer, equipped with a (284$\times$423)\,$\upmu$m$^2$ W-TES of 200\,nm thickness, which has 
two (526$\times$1027)\,$\upmu$m$^2$ phonon collectors (Al/W bilayers) of 1\,$\upmu$m thickness~\cite{Angloher:2004tr}.~
It is mounted on the lid of the copper holder, which is used to protect the NaI from humid air. The light detector is irradiated with a second $^{55}$Fe calibration source of similar activity as the one shining on the absorber. A picture of the light detector (Fig.~\ref{fig:LDholder}) and an enlarged view of the wire bonding of its TES (Fig.~\ref{fig:LDtes}) are shown in Fig.~\ref{fig:LD}. In the following, we refer to the NaI-remoTES as the phonon channel and the light detector as the light channel, interchangeably. The description of the detector components is summarised in Tab.~\ref{tab:dimensions}.

\begin{table}[!htb]
    \centering
    \begin{tabular}{l||l}
    \hline
        Component~&  ~Dimension\\ \hline\hline
         NaI-absorber~           &  ~(10$\times$10$\times$10) mm$^3$ \\ 
         Au-pad on NaI~          &  ~Area: 4 mm$^2$ \\
                                 &  ~Thickness: 1 $\upmu$m \\
                                 &  ~Glue: EPO-TEK 301-2 \\ 
                                 &  ~RRR: $\approx 22$ \\ 
         Two Au-wires~           &  ~Lengths: (6.7, 10.3) mm\\
                                 &  ~Diameter: 17 $\upmu$m \\ \hline
        Al$_2$O$_3$ wafer~       &  ~(10$\times$20$\times$1) mm$^3$ \\ 
        W-TES on wafer~          &  ~Area: (100$\times$400)~$\upmu$m$^2$ \\
                                 &  ~Thickness: 156 nm \\ 
                                 &  ~Area heater: (200$\times$150)~$\upmu$m$^2$ \\
                                 &  ~Thickness heater: 100 nm \\ \hline
        SOS wafer         &  ~(20$\times$20$\times$0.4)~mm$^3$ \\ 
        W-TES on SOS wafer  &  ~Area: (300$\times$450) $\upmu$m$^2$ \\
                                 &  ~Thickness: 200 nm \\ \hline
        
    \end{tabular}
    \caption{Dimensions of all components of phonon detector and light detector.}
    \label{tab:dimensions}
\end{table}

\section{Data taking}
\label{sec:datataking}
The measurement was carried out in an above-ground wet dilution refrigerator at the Max Planck Institute for Physics in Munich. The cryostat is equipped with four superconducting quantum interference devices (SQUIDs) from the APS company~\cite{APS} and continuously read out with a 16-bit analog-digital converter at a sampling frequency of \unit[50]{kS/s}.
To reduce the background event rate induced by cosmic and ambient radiation, a lead wall with a thickness of 10\,cm was built around the refrigerator. Three subsequent datasets were recorded and are discussed in this work: a background dataset with only the (module-internal) $^{55}$Fe sources present, a $^{57}$Co calibration dataset, and an additional neutron calibration dataset using an AmBe source. The duration and effective exposure of each dataset are given in Table \ref{tab:filelist}.

\begin{table*}[!htb]
    \centering
    \begin{tabular}{cccc}
        \hline
        Measuring time (h) & Calibration sources & Event rate (cps) & Exposure \unit[]({g$\cdot$d})\\ \hline \hline
        17.73           & $^{57}$Co           & 0.57      &   2.73       \\ 
        16.40           & AmBe                  & 0.62      &   2.53    \\
        24.61           & (none)                 & 0.37      &   3.79    \\ \hline
    \end{tabular}
    \caption{Measuring times, exposures, event rates and calibration sources for the three datasets.}
    \label{tab:filelist}
\end{table*}

\section{Data analysis}
\label{sec:dataanalysis}
\begin{figure}[ht]
\begin{subfigure}[b!]{\linewidth}
    \includegraphics[width=0.9\linewidth]{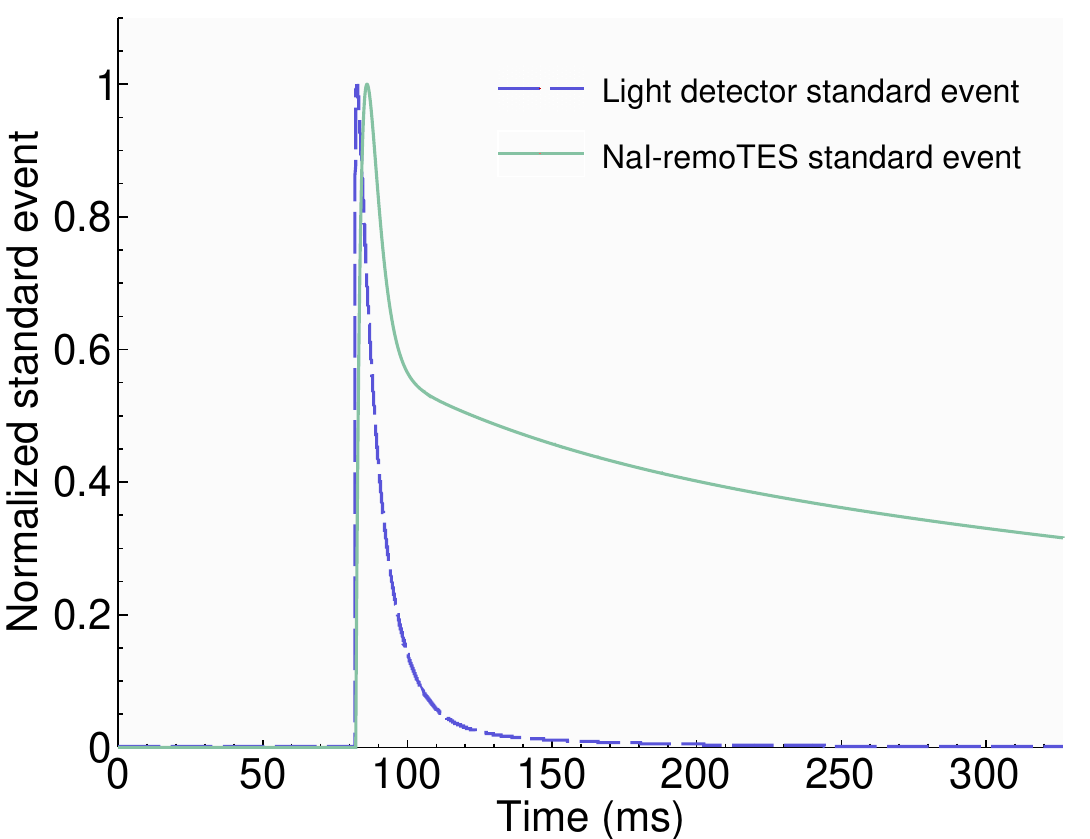}
    \caption{}
    \label{fig:sev}
\end{subfigure}\hfill%
\begin{subfigure}[b!]{\linewidth}
     \includegraphics[width=0.9\linewidth]{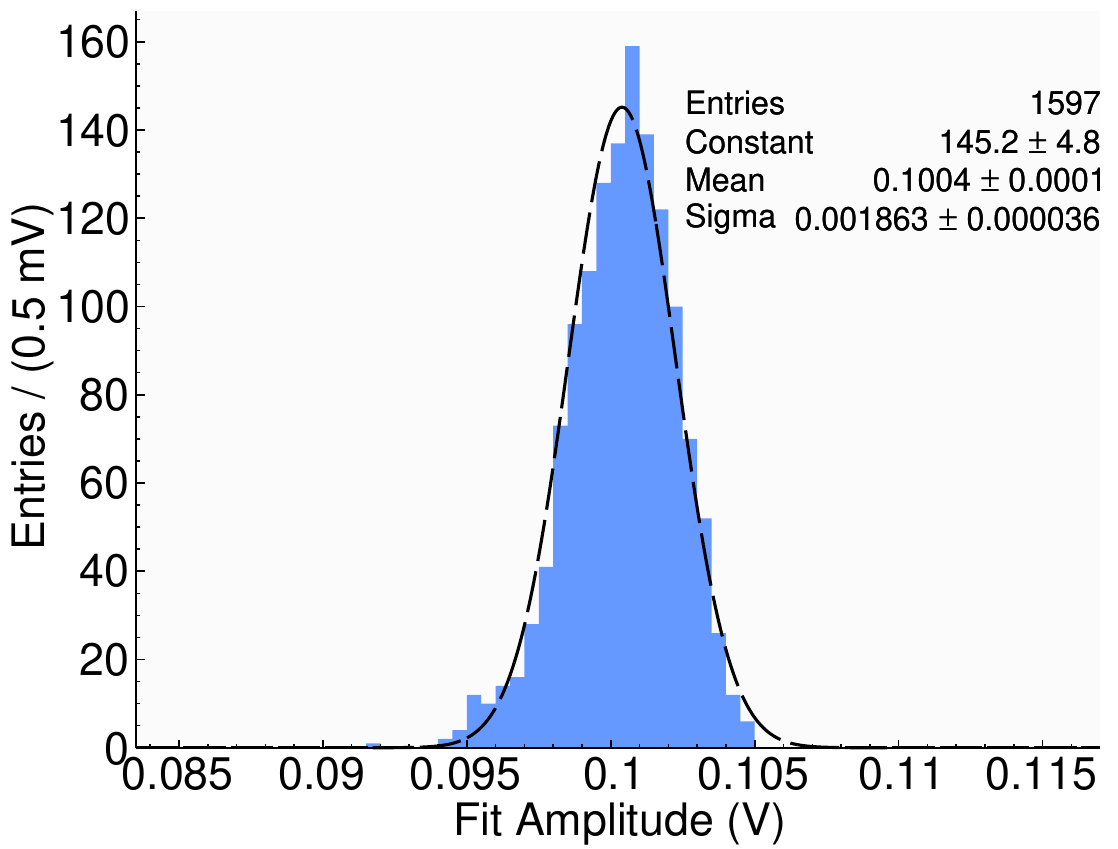}
\caption{}
\label{fig:sim}
\end{subfigure}
    \caption{(a) Normalized standard events in the light detector (dashed, dark-blue curve) and in the NaI-remoTES detector (solid, water-green curve). (b) Reconstructed pulse height (truncated fit) distribution for artificial pulses, which are obtained by superimposing an averaged signal pulse onto empty traces for the phonon channel.}
\label{fig:sev_and_resolution}
\end{figure}

\begin{figure*}[!htb]
    \centering
    \includegraphics[width=0.9\linewidth]{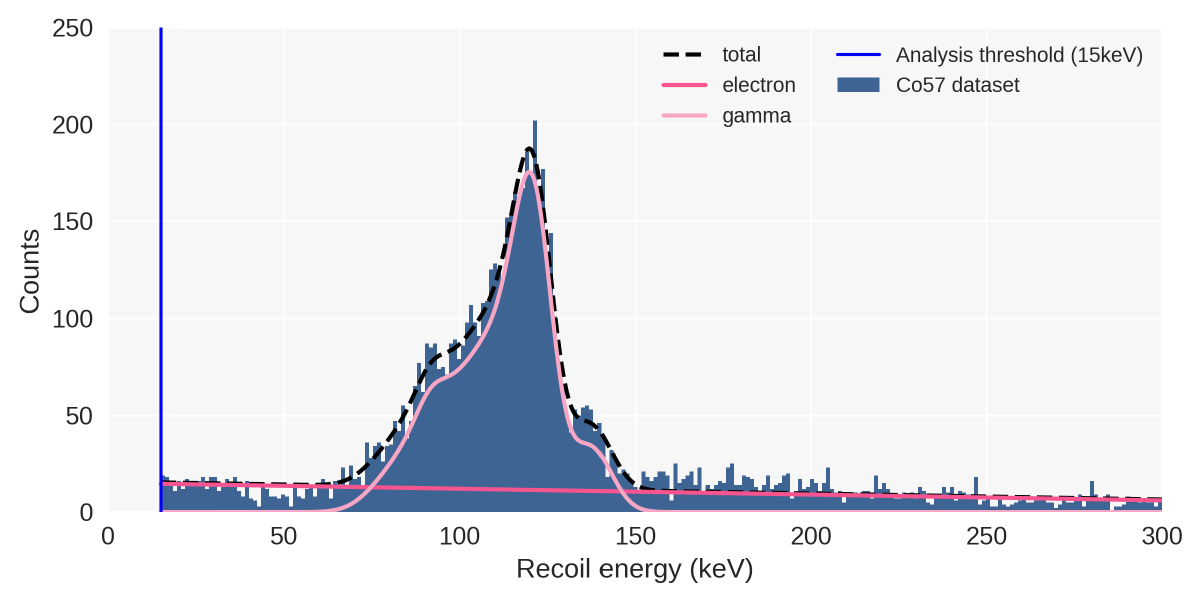}
    \caption{Energy spectrum from the $^{57}$Co $\gamma$ calibration measurement. Different components are fitted using a parametric description. The fit includes gaussian peaks at \unit[136]{keV} and \unit[122]{keV} on top of a linearly decreasing background, as well as a $\gamma$ escape peak due to I K-$\alpha$, expected at around \unit[89]{keV}. The parametric fit also considers `shoulders' on the left side of each peak, which originate from $^{57}$Co $\gamma$s depositing a fraction of their energy in parts of the setup before reaching the detector.}
    \label{fig:spectrum}
\end{figure*}

For the three datasets discussed above, the continuous stream from the phonon channel is triggered in software using an optimal filter trigger (cf. Ref.~\cite{gatti_processing_1986} and Ref.~\cite{angloher_results_2017}). The light channel is read out in parallel. The filter is created from a parametric description of the NaI channel pulse shape based on Ref.~\cite{probst_model_1995} and a noise power spectrum obtained from randomly drawn empty noise traces. The typical pulse shape of absorber recoils for the NaI channel is shown in Fig.~\ref{fig:sev} and features a very long decay time. A baseline energy resolution of the phonon channel of \unit[2.07$\pm0.06$]{keV} is determined by superimposing the pulse shape onto a set of randomly drawn empty baselines and reconstructing these artificial events. The resulting amplitude distribution is illustrated in Fig.~\ref{fig:sim}. The baseline resolution of the SOS light detector was determined to be \unit[2.02$\pm0.05$]{keV$_{ee}$} (electron-equivalent) using the same technique.\\The datasets were triggered with a threshold of \unit[10]{keV}, where no noise triggers were observed.  
The energy scale for the aforementioned results comes from a fit to the peaks visible in the $^{57}$Co dataset; this is illustrated in Fig.~\ref{fig:spectrum}. Peaks from the $^{55}$Fe source could not be observed, as their energies of \unit[5.9]{keV} and \unit[6.4]{keV} are below the energy threshold. The optimum filter amplitude is used as an energy estimator. In the energy range from threshold up to $\sim$\unit[500]{keV}, the detector response is in good approximation linear.
All datasets were cleaned by applying a set of quality cuts. Severely unstable detector operation intervals are removed by monitoring the reconstructed amplitude of injected heater pulses over time. Single voltage spike events are removed by cutting on the ratio of the numerical derivative of the pulse to the baseline RMS. The RMS from the optimal filter reconstruction and the RMS from an additional truncated standard event fit reconstruction (cf. e.g. Ref.~\cite{angloher2009commissioning}) are used to remove pile-up events and events from particle interactions in the sapphire wafer. The effect of each quality cut is assessed by simulating pulses with a flat energy spectrum on the set of randomly drawn detector baselines and studying the fraction of surviving events as a function of the simulated energy. We find that the detector threshold of \unit[10]{keV} is only nominal, i.e. no simulated signal event survives down to this energy. This is due to varying noise conditions in the above-ground setup, very long pulse decay time, and the presence of particle recoils in the wafer, which require strong quality cuts to be discarded. An analysis threshold of \unit[15]{keV} is used in the following, where the signal survival probability is still about \unit[5]{\%}. Above this threshold, no wafer-induced events or noise events are observed in the background dataset.

\begin{figure*}[ht]
    \centering
    \includegraphics[width=0.9\linewidth]{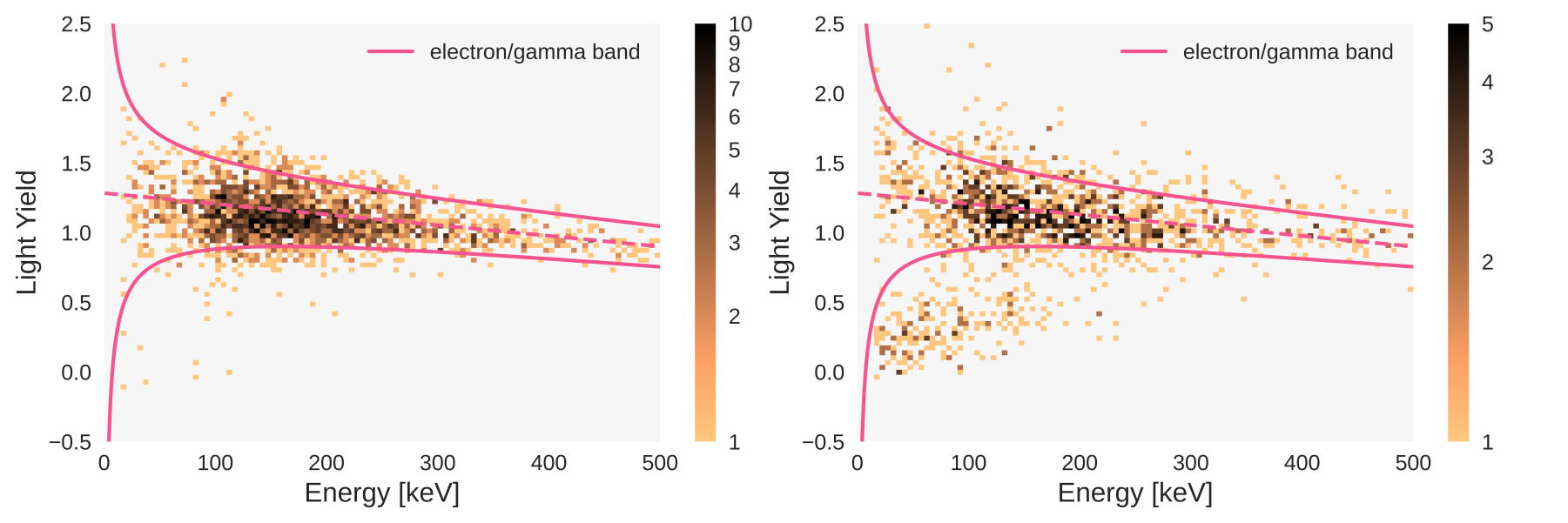}
    \caption{2D histograms of LY versus phonon channel energy with color-coded number of entries. Left: background dataset with parametric description of the \ega band obtained from the log-likelihood fit. Right: same plot for the AmBe neutron calibration dataset. A second population of events is visible below the \ega band.}
    \label{fig:bandfit}
\end{figure*}

\section{Discrimination of Nuclear Recoil Bands}
\label{sec:results}

The light yield (LY) is defined as the ratio of the energy measured in the light channel and the energy measured in the phonon channel for each event. It enables the discrimination of different recoil event classes. In Fig.~\ref{fig:bandfit} the LY is shown as a function of the phonon channel energy for the background and the neutron calibration datasets, respectively. We use a parametrization of the recoil bands and of the spectra which contribute to them, described in detail in~\cite{angloher2022testing}. For events in the electron band, the LY is normalized to 1. Two quenched bands are expected for Na and I nuclear recoils due to the different masses of the nuclei. We observe that these cannot be reliably separated by the likelihood fit, which yields a high correlation between parameters for the two bands. Therefore, we conservatively estimate the parameters of the \ega band from the background dataset, and compare them to the observed events in the neutron calibration dataset. We will address a modification of the band description and separation of the individual nuclear recoil bands in a future publication. Figure~\ref{fig:bandfit} shows the result of the likelihood bandfit to the background dataset (left panel) and to the neutron calibration dataset (right panel), using \unit[90]{\%} boundaries for the \ega band. It can be clearly seen that a new population of events appears below the \ega band, far outside the \unit[2]{$\sigma$} boundary. The additional band is quenched in LY to $\sim0.5$ at energies above \unit[100]{keV}, and the LY appears to be decreasing as the recoil energy approaches the threshold. In Ref.~\cite{cintas2021quenching}, a similar decline of the quenching factor in NaI at lower energies was reported for measurements at room temperature and with Tl-doped crystals. Above ca.~\unit[100]{keV}, the quenching factor observed with our prototype is even higher than in Ref.~\cite{cintas2021quenching}. At increasing energy, a downward tilt of the \ega band is visible, which is due to increasing nonlinearity in the light detector response.\\

\section{Conclusions}
\label{sec:conclusion}
This measurement marks the first proof of event-by-event particle discrimination in a cryogenic NaI detector. We operated a COSINUS prototype with a remoTES sensor, which displayed a baseline resolution of \unit[2]{keV} despite suboptimal, above-ground conditions. It was calibrated with a $^{57}$Co $\gamma$ source and analysed with a nuclear recoil threshold of \unit[15]{keV}. Particle discrimination was verified with neutrons from an AmBe source. In Ref.~\cite{angloher2023first}, the remoTES design was suggested as an improved readout for delicate absorber materials, which are for example hygroscopic or feature a low melting point. This work shows that the design is indeed suitable for NaI absorbers.  
The next step in the COSINUS detector optimization strategy in the direction of achieving an energy threshold of \unit[1]{keV} is an underground measurement with a similar detector, in order to assess its performance in a low-background environment. This measurement has already been performed, and will be subject of a future publication.

\section*{Acknowledgments} 
This measurement was possible thanks to the use of the CRESST cryogenic facility and detector production infrastructure at the Max Planck Institute for Physics (MPP). Similarly, the support from the MPP mechanical workshop was invaluable. We thank SICCAS for producing the NaI crystal employed in this measurement. This work has been supported by the Austrian Science Fund FWF, stand-alone project AnaCONDa [P 33026-N]. This project was supported by the Austrian Research Promotion Agency (FFG), project ML4CPD.

\bibliography{all,remoTES}

\begin{thebibliography}{15}
\providecommand{\natexlab}[1]{#1}
\providecommand{\url}[1]{\texttt{#1}}
\expandafter\ifx\csname urlstyle\endcsname\relax
  \providecommand{\doi}[1]{doi: #1}\else
  \providecommand{\doi}{doi: \begingroup \urlstyle{rm}\Url}\fi

\bibitem[Billard et~al.(2022)]{billard2022direct}
J.~Billard et~al.
\newblock Direct detection of dark matter—{APPEC} committee report.
\newblock \emph{Reports on Progress in Physics}, 85\penalty0 (5):\penalty0
  056201, 2022.
\newblock \doi{10.1088/1361-6633/ac5754}.

\bibitem[Bernabei et~al.(2022)]{bernabei2022recent}
R.~Bernabei et~al.
\newblock Recent results from {DAMA/LIBRA} and comparisons.
\newblock \emph{Moscow University Physics Bulletin}, 77\penalty0 (2):\penalty0
  291--300, 2022.
\newblock \doi{10.15407/jnpae2021.04.329}.

\bibitem[Adhikari et~al.(2022)]{adhikari2022three}
G.~Adhikari et~al.
\newblock Three-year annual modulation search with {COSINE}-100.
\newblock \emph{Physical Review D}, 106\penalty0 (5):\penalty0 052005, 2022.

\bibitem[Amar{\'e} et~al.(2022)]{amare2022dark}
J.~Amar{\'e} et~al.
\newblock Dark matter annual modulation with {ANAIS}-112: {T}hree years
  results.
\newblock \emph{Moscow University Physics Bulletin}, 77\penalty0 (2):\penalty0
  322--326, 2022.

\bibitem[Angloher et~al.(2017)]{angloher2017results}
G.~Angloher et~al.
\newblock Results from the first cryogenic {N}a{I} detector for the {COSINUS}
  project.
\newblock \emph{Journal of Instrumentation}, 12\penalty0 (11):\penalty0 P11007,
  2017.
\newblock \doi{10.1088/1748-0221/12/11/P11007}.

\bibitem[Angloher et~al.()]{angloher2023first}
G.~Angloher et~al.
\newblock First measurements of remo{TES} cryogenic calorimeters:
  {E}asy-to-fabricate particle detectors for a wide choice of target materials.
\newblock \emph{Nuclear Instruments and Methods in Physics Research Section A:
  Accelerators, Spectrometers, Detectors and Associated Equipment}.
\newblock \doi{10.1016/j.nima.2022.167532}.

\bibitem[Note1()]{Note1}
Note1.
\newblock https://www.epotek.com/.

\bibitem[Angloher et~al.(2005)]{Angloher:2004tr}
G.~Angloher et~al.
\newblock {Limits on WIMP dark matter using scintillating CaWO$_4$ cryogenic
  detectors with active background suppression}.
\newblock \emph{Astropart. Phys.}, 23:\penalty0 325--339, 2005.
\newblock \doi{10.1016/j.astropartphys.2005.01.006}.

\bibitem[APS()]{APS}
{Applied Physics Systems}.
\newblock \url{https://appliedphysics.com/}.

\bibitem[Gatti and Manfredi(1986)]{gatti_processing_1986}
E.~Gatti and P.~F. Manfredi.
\newblock Processing the signals from solid-state detectors in
  elementary-particle physics.
\newblock \emph{Riv. Nuovo Cim.}, 9\penalty0 (1):\penalty0 1--146, Jan. 1986.
\newblock ISSN 1826-9850.
\newblock \doi{10.1007/BF02822156}.

\bibitem[Angloher et~al.(2017)]{angloher_results_2017}
G.~Angloher et~al.
\newblock Results on {MeV}-scale dark matter from a gram-scale cryogenic
  calorimeter operated above ground.
\newblock \emph{Eur. Phys. J. C}, 77\penalty0 (9):\penalty0 637, Sept. 2017.
\newblock ISSN 1434-6044, 1434-6052.
\newblock \doi{10.1140/epjc/s10052-017-5223-9}.

\bibitem[Pr{\"o}bst et~al.(1995)]{probst_model_1995}
F.~Pr{\"o}bst et~al.
\newblock Model for cryogenic particle detectors with superconducting phase
  transition thermometers.
\newblock \emph{Journal of low temperature physics}, 100:\penalty0 69--104,
  1995.
\newblock \doi{10.1007/BF00753837}.

\bibitem[Angloher et~al.(2009)]{angloher2009commissioning}
G.~Angloher et~al.
\newblock Commissioning run of the {CRESST-II} dark matter search.
\newblock \emph{Astroparticle Physics}, 31\penalty0 (4):\penalty0 270--276,
  2009.
\newblock \doi{10.1016/j.astropartphys.2009.02.007}.

\bibitem[Angloher et~al.(2022)]{angloher2022testing}
G.~Angloher et~al.
\newblock {Testing spin-dependent dark matter interactions with lithium
  aluminate targets in CRESST-III}.
\newblock \emph{Physical Review D}, 106\penalty0 (9):\penalty0 092008, 2022.
\newblock \doi{10.1103/PhysRevD.106.092008}.

\bibitem[Cintas et~al.(2021)]{cintas2021quenching}
D.~Cintas et~al.
\newblock {Quenching Factor consistency across several NaI (Tl) crystals}.
\newblock In \emph{Journal of Physics: Conference Series}, volume 2156, page
  012065. IOP Publishing, 2021.
\newblock \doi{10.1088/1742-6596/2156/1/012065}.

\end{thebibliography}

\end{document}